\documentclass[preprint,5p]{elsarticle}

\usepackage{epsfig,latexsym,amssymb,amsmath}
\usepackage{graphicx}
\usepackage{epstopdf}
\usepackage{color}
\usepackage{dcolumn}
\usepackage{bm}
\usepackage{hyperref}
\usepackage{ulem}

\def\g{\gamma}

\def\s{\sigma}

\def\x{\xi}

\def\be{\begin{eqnarray}}
\def\ee{\end{eqnarray}}
\def\bea{\begin{eqnarray}}
\def\eea{\end{eqnarray}}

\def\bbuildrel#1_#2^#3{\mathrel{\mathop{\kern 0pt#1}\limits_{#2}^{#3}}}
\def\slash#1{\setbox0=\hbox{$#1$}#1\hskip-\wd0\dimen0=5pt\advance
       \dimen0 by-\ht0\advance\dimen0 by\dp0\lower0.5\dimen0\hbox
         to\wd0{\hss\sl/\/\hss}}

\newcommand{\gae}{\lower 2pt \hbox{$\, \buildrel {\scriptstyle >}\over {\scriptstyle
\sim}\,$}}
\newcommand{\lae}{\lower 2pt \hbox{$\, \buildrel {\scriptstyle <}\over {\scriptstyle
\sim}\,$}}

\newcommand{\beq}{\begin{eqnarray}}
\newcommand{\eeq}{\end{eqnarray}}

\newcommand{\ba}{\begin{array}}
\newcommand{\ea}{\end{array}}

\long\def\symbolfootnote[#1]#2{\begingroup%
\def\thefootnote{\fnsymbol{footnote}}\footnote[#1]{#2}\endgroup}

\def\lsim{\mathrel{\rlap{\lower4pt\hbox{\hskip1pt$\sim$}}
    \raise1pt\hbox{$<$}}}         
\def\gsim{\mathrel{\rlap{\lower4pt\hbox{\hskip1pt$\sim$}}
    \raise1pt\hbox{$>$}}}         

\def\lsim{\:\raisebox{-0.5ex}{$\stackrel{\textstyle<}{\sim}$}\:}
\def\gsim{\:\raisebox{-0.5ex}{$\stackrel{\textstyle>}{\sim}$}\:}

\def\issue(#1,#2,#3){{\bf #1}, #2 (#3)}
\def\opcit(#1){ {\em op. cit.}, #1}

\def\APP(#1,#2,#3){Acta Phys.\ Polon.\ \issue(#1,#2,#3)}
\def\ARNPS(#1,#2,#3){Ann.\ Rev.\ Nucl.\ Part.\ Sci.\ \issue(#1,#2,#3)}
\def\CPC(#1,#2,#3){Comp.\ Phys.\ Comm.\ \issue(#1,#2,#3)}
\def\CIP(#1,#2,#3){Comput.\ Phys.\ \issue(#1,#2,#3)}
\def\EPJC(#1,#2,#3){Eur.\ Phys.\ J.\ C\ \issue(#1,#2,#3)}
\def\EPJD(#1,#2,#3){Eur.\ Phys.\ J. Direct\ C\ \issue(#1,#2,#3)}
\def\IEEETNS(#1,#2,#3){IEEE Trans.\ Nucl.\ Sci.\ \issue(#1,#2,#3)}
\def\IJMP(#1,#2,#3){Int.\ J.\ Mod.\ Phys. \issue(#1,#2,#3)}
\def\JHEP(#1,#2,#3){J.\ High Energy Physics \issue(#1,#2,#3)}
\def\JPG(#1,#2,#3){J.\ Phys.\ G \issue(#1,#2,#3)}
\def\MPL(#1,#2,#3){Mod.\ Phys.\ Lett.\ \issue(#1,#2,#3)}
\def\NP(#1,#2,#3){Nucl.\ Phys.\ \issue(#1,#2,#3)}
\def\NIM(#1,#2,#3){Nucl.\ Instrum.\ Meth.\ \issue(#1,#2,#3)}
\def\PL(#1,#2,#3){Phys.\ Lett.\ \issue(#1,#2,#3)}
\def\PRD(#1,#2,#3){Phys.\ Rev.\ D \issue(#1,#2,#3)}
\def\PRL(#1,#2,#3){Phys.\ Rev.\ Lett.\ \issue(#1,#2,#3)}
\def\SJNP(#1,#2,#3){Sov.\ J. Nucl.\ Phys.\ \issue(#1,#2,#3)}
\def\ZPC(#1,#2,#3){Zeit.\ Phys.\ C \issue(#1,#2,#3)}

\def\beq{\begin{eqnarray}}
\def\eeq{\end{eqnarray}}
\def\bea{\begin{eqnarray}}
\def\eea{\end{eqnarray}}
\def\to{\rightarrow}

\journal{Physics Letters B}
\begin{document}
\DeclareGraphicsExtensions{.jpg,.pdf,.mps,.png,}

\begin{frontmatter}

\title{750 GeV Resonance in the Gauged $U(1)'$-Extended MSSM}

\author[nbi]{Yun Jiang}
\ead{yunjiang@nbi.ku.dk}
\author[hkust]{Ying-Ying Li}
\ead{ylict@connect.ust.hk}
\author[hkust]{Tao Liu}
\ead{taoliu@ust.hk}

\address[nbi]{NBIA and Discovery Center, Niels Bohr Institute, University of Copenhagen,\\ Blegdamsvej 17, DK-2100, Copenhagen, Denmark}
\address[hkust]{Department of Physics, The Hong Kong University of Science and Technology, \\Clear Water Bay, Kowloon, Hong Kong S.A.R., P.R.C.}

\begin{abstract}
Recently the ATLAS and CMS collaborations at the LHC announced their observation of a potential 750 GeV di-photon resonance, after analyzing the $\sqrt{s}=13$~TeV LHC data. This observation has significant implications for low-energy supersymmetry. Beyond the MSSM and the NMSSM,
we study the MSSM-extensions with an extra $U(1)'$ gauge symmetry. The anomaly cancellation and the spontaneous breaking of the non-decoupled $U(1)'$ generally require introducing vector-like supermultiplets (both colored and color-neutral ones) and singlet supermultiplets, respectively. We illustrate that the potential 750 GeV resonance ($Y$) can be accommodated in various mechanisms, as a singlet-like scalar or pseudoscalar. Three benchmark scenarios are presented: (1) vector-like quarks (VLQ) mediated $pp \to Y \to \gamma \gamma$; (2) scalar VLQ mediated $pp \to Y \to \gamma \gamma$;  (3) heavy scalar (pseudo-scalar) $H/A$ associated production $pp \to H^*/A^* \to Y H/h$. Additionally, we notice that the $Z'$-mediated vector boson fusion production and $Z'$-associated production $pp \to Y qq'$, if yielding a signal rate of the observed level, might have been excluded by the searches for $Z'$ via Drell-Yan process at the LHC.

\end{abstract}


\end{frontmatter}

\section{Introduction}

Recently the ATLAS \cite{ATLAS:2015} and CMS \cite{CMS:2015} collaborations at the Large Hadron Collider (LHC) reported their first results on the di-photon analysis at energy $\sqrt{s}=13$~TeV. Both collaborations have found an excess over the background at the di-photon ($\gamma\gamma$) invariant mass $m_{\g\g}\sim750$ GeV, suggesting a signal cross section of $10\pm 3$ fb at the ATLAS and $6\pm 3$ fb at the CMS. The local significance of this signal is modest, about $3.9 \sigma$ and $2.6 \sigma$ at the ATLAS and CMS, respectively. Meanwhile, no notable excess was observed in any other channels, including the $t\bar t,~hh,~WW,~ZZ$ and di-jet final states. Also, this resonance was not seen in the 8 TeV Run-1 data, although both the ATLAS and CMS presented a mild upward protuberance in their di-photon analyses. 

This observation might indicate a new breakthrough achieved at the LHC after the discovery of the 125 GeV standard model (SM)-like Higgs boson, though more measurements and statistics are needed for us to make a conclusive statement. If it turns out to be real, this observation may have deep implications for new physics. So far, many scenarios have been proposed for interpreting this di-photon excess either in concrete models or by means of model-independent EFT approach. In either case, new particles such as vector-like fermions~\cite{diphoton750smVLQ,Angelescu:2015uiz,Gupta:2015zzs,diphoton7502hdmVLQ,diphoton750nmssmVLQ}, hypercharged scalars~\cite{diphoton750scalarVLQ} or exotic vector boson~\cite{deBlas:2015hlv} are introduced to induce the $Ygg$ and/or $Y\g\g$ loops for the 750 GeV resonance $Y$, which has null or very small couplings with the SM particles. This idea has also been implemented in the context of non-supersymmetric extensions of the SM with an extra $U(1)'$ or enlarged gauge symmetries~\cite{diphoton750exsym}. On the other hand, recent studies also discuss an axion-like explanation of the di-photon excess~\cite{diphoton750axion}, in the context  of composite models~\cite{diphoton750comp}. Beyond the 4D models, Ref.~\cite{diphoton750xdim} considers the resonance as a radion from the extra dimension at the conformal limit. In addition, one has extensively studied the connection of this potential 750~GeV resonance to dark matter~\cite{diphoton750dm}, neutrino physics~\cite{diphoton750neutrino} or more fundamental theories~\cite{diphoton750string}. Other than the direct production of di-photon from a heavy resonance at 750 GeV,  exotic possibilities that the di-photon signal arises from a three-body decay~\cite{diphoton7503body} along with an undetectable particle or from two bunches of photon jets emitted by a pair of pseudoscalars~\cite{diphoton7504body} have also been explored. 

This naturally raises the question if this resonance can be accommodated in supersymmetric theories because of their popularity in particle physics. This topical issue has been partially addressed in Ref.~\cite{diphoton750susy}. In the Minimal Supersymmetric Standard Model (MSSM), the candidates for this resonance could be the heavy CP-even Higgs $H$ or the CP-odd Higgs $A$ as the lightest CP-even Higgs $h$ must be at 125 GeV. However, it has been shown in~\cite{Angelescu:2015uiz,Gupta:2015zzs,DiChiara:2015vdm} that having the $H$ or the $A$ at 750 GeV, at most, approaches the cross section in the di-photon final state via the gluon-fusion production, $\sigma(gg \to H/A) {\rm BR} (H/A \to \gamma\gamma) \sim 10^{-2}$ fb, which is three orders smaller than the observed level at 13~TeV LHC. The reasons are in the following. First, the contribution of the charged Higgs to the loop-induced $H\gamma\gamma$ coupling is negligible as $m_{H^\pm} \simeq m_H \simeq m_A \simeq 750$~GeV yields a small coupling $g_{H H^+H^-} \simeq \mathcal{O}(1)$. Second, relatively light stop quark could have made significant contributions to $H/A \to \gamma\gamma$ and $gg\to H/A$, given a small $\tan\beta$. This however makes it difficult to accomplish $m_h =125$~GeV, because of the tree-level prediction $(m_h)_{\rm tr} \approx m_Z \cos 2\beta$ in the MSSM (actually to acommondate such a SM-like Higgs boson, careful choice of parameters must be made~\cite{Carena:2011aa,Christensen:2012ei,Draper:2011aa,Cao:2012fz,Feng:2012jf,Hagiwara:2012mg,Christensen:2012si,Arbey:2012dq,Benbrik:2012rm}).  In the next-to-MSSM (NMSSM), the strong tension between the $m_h$ and $\tan\beta$ is slightly relaxed. But it is still difficult to accommodate a 750 GeV resonance with the observed properties. The reason is that supersymmetry may impose large corrections to the $h$ couplings, spoiling their SM-like properties, if it plays a non-trivial role in defining the properties of the 750 GeV resonance. As a result, it is well-motivated to consider the MSSM extensions which can (1) give additional contributions to the mass of the SM-like Higgs boson at tree level~\cite{Benbrik:2012rm,Vasquez:2012hn,Ellwanger:2011aa,Hall:2011aa} either via new F-terms or via new D-terms or via both in the Higgs potential, and (2) introduce new candidates for the 750 GeV resonance, without spoiling the properties of the SM-like Higgs boson.   

To this end, we consider the possibility of the MSSM extension with an additional gauge symmetry $U(1)'$, with R-parity conserved and no CP-violation. Such a gauge symmetry extensively exists in supersymmetric theories, e.g., the GUT (for a review, see, e.g.,~\cite{Langacker:2008yv}). In this paper we focus on a gauged Peccei-Quinn (PQ) symmetry, under which the Higgs fields carry non-trivial charges by definition~\cite{Peccei:1977hh}. If its D-term is non-decoupling, the additional gauge symmetry can provide non-trivial contribution to the tree-level mass of the SM-like Higgs boson~\cite{Batra:2003nj}. Additionally, an effective $\mu$ parameter could be dynamically generated in this framework~\cite{Kim:1983dt,Kim:2012at}. It is known that the bare $\mu$ term is forbidden by the $U(1)'$ gauge symmetry. Whereas, an effective one can be produced via ${\bf W_H} \sim \lambda {\bf S H_u H_d}$, yielding $\mu_{\rm eff} = \lambda v_S$. Here ${\bf H_d}$ and ${\bf H_u}$ are down- and up-type Higgs supermultiplets, and ${\bf S}$ is a SM-singlet supermultiplet. In this context, the potentially existing 750 GeV resonance $Y$ can be interpreted as a singlet-like scalar or a singlet-like pseudoscalar. 

There are three classes of couplings which are relevant to our studies below.
\begin{itemize}
\item The resonance couplings with vector-like fermions $\sim Y F_V F_V^c$, and their superpartners $\sim Y \tilde F_V \tilde F_V^*, Y \tilde F_V^c \tilde F_V^{c*}$. Here $F_V$ and $F_V^c$ are  ``vector-like'' with respect to the SM gauge symmetries. They could be chiral under the $U(1)'$ gauge symmetry. It is well-known that the anomaly cancellation for the whole set of the gauge symmetries generically requires the introduction of both colored and color-neutral vector-like supermultiplets, unless some Chern-Simons term is properly incorporated. Because the singlet supermultiplets carry a $U(1)'$ charge as well, they may couple with these vector-like supermultiplets, yielding a mass for these exotics. 

\item The resonance couplings with the MSSM-like Higgs bosons $\sim Y H H, YA h$. Here $H/A$ are the MSSM-like neutral Higgs bosons.

\item The resonance couplings with the $U(1)'$ gauge boson $\sim Y Z' Z'$, since the singlet supermultiplet carries a $U(1)'$ gauge charge. 

\end{itemize}
In the presence of these couplings, $Y$ could be produced with a relatively large boost for the cross section at 13 TeV LHC, compared to that at the 8 TeV LHC. The main production mechanisms include: 
\begin{enumerate}
 
 \item Vector-like quark (VLQ) mediated gluon fusion. $Y$ is singlet-like, with its coupling to the SM fermions and their superpartners suppressed by a small mixing angle. But its coupling with VLQ can yield a large effective coupling with di-gluon (as well as di-photon), resulting in a boost factor as large as $4.7$.
 
\item Scalar VLQ (SLVQ) mediated gluon fusion. 
 
 \item $H/A$-mediated production. Compared to the first two mechanisms, a relatively large boost factor arises due to the increased energy threshold. 
 
 \item $Z'$-mediated vector boson fusion and $Z'$-associated production. Compared to the other mechanisms, these ones are constrained by the current experimental bounds much more tightly, unless the leptonic decay of $Z'$ can be greatly suppressed. 
 
 \end{enumerate}
With these production mechanisms, the 750 GeV resonance can be explained, given a decay branching ratio Br$(Y\to \gamma \gamma)$ of subpercent level or above. This could be achieved with its effective di-photon coupling mediated by VLQ (SVLQ) and VLL (SVLL).

We organize the paper as follows. In Section II, we present a general discussion on the MSSM extension with a PQ-type $U(1)'$ gauge symmetry, as well as the properties of the candidate for the 750 GeV resonance. In Section III, we identify three benchmark scenarios where the experimental data can be properly fit, using an anomaly-free model for illustration. Section IV contains our conclusions.

\section{The Gauged $U(1)'$-Extended MSSM: General Discussions}

In the gauged $U(1)'$-extended MSSM, a list of chiral supermultiplets in the Higgs sector typically includes (see, e.g.,~\cite{Langacker:2008yv})
$$
\{\bf H_a\} = \{ \bf H_d, H_u, S, S_i \}, \ \ i = 1, 2, 3, ...
$$ 
Here ${ \bf S}$ and ${\bf S_i }$ are the SM singlets, both of which carry $U(1)'$ gauge charge.  By coupling with ${ \bf H_d}$ and ${\bf H_u}$, ${ \bf S}$ can dynamically  generate an effective MSSM $\mu$ parameter. ${ \bf S_i}$  is introduced for anomaly cancellation. The $U(1)'$ breaking scale is defined as
\beq
f_{Z'} &=& \left(\sum_a q_a^2 v_i^2\right)^{1/2}  \\ &=& \left(q_S^2 v_S^2 + q_{H_d}^2 v_d^2 + q_{H_u}^2 v_u^2 + \sum_{i} q_{S_i}^2 v_{S_i}^2 \right)^{1/2}. \nonumber
\eeq
Here $q_a$ and $v_a$ denote the $U(1)'$ charge carried by $H_a$, and its VEV, respectively.   

Then we can write down the tree-level $Z-Z'$ mass matrix 
\begin{eqnarray}
\mathcal{M}_{Z-Z'} =\left(\begin{matrix}  
M_{Z}^2 & M_{Z Z'}^2 \cr
M_{Z Z'}^2 &  M_{Z'}^2  \end{matrix} \right) ~,~ \,
\end{eqnarray}
with
\begin{eqnarray}
M_{Z}^2 = {G^2\over 2} (v_d^2 +v_u^2)
~,~ M_{ Z'}^2 = 2 g_{Z'}^2  f_{Z'}^2 
\end{eqnarray} 
\begin{eqnarray}
M_{Z Z'}^2 = g_{Z'} G ( q_{H_d} v_d^2 - q_{H_u} v_u^2)
~.~\,
\end{eqnarray} 
Here $G=\sqrt{g_1^2 + g_2^2}$ is a combination of $SU(2)_L \times U(1)_Y$ gauge couplings, and $g_{Z'}$ is the $U(1)'$ gauge coupling. This leads to a $Z-Z'$ mixing angle 
\begin{eqnarray}
\alpha_{Z-Z'} = {1\over 2} {\rm arctan} \left({{2 M_{ZZ'}^2}
\over\displaystyle {M_{Z'}^2 - M_Z^2}}\right) \sim  \frac{G}{g_{Z'}} \frac{v^2}{f_{Z'}^2}  
~,~\,
\end{eqnarray} 
which is constrained to be of or less than $\mathcal O(10^{-3})$ by EW precision tests~\cite{Erler:2009jh}. Given $\lambda v_S \sim v = \sqrt{v_d^2 + v_u^2}$=174~GeV, we expect a large contribution to $f_{Z'}$ arise from $v_{S_i} \gg v, v_S $. In this case, the set of $\{\bf S_i\}$ 
define a so-called secluded $U(1)'$-breaking sector~\cite{Erler:2002pr}: they mainly serve as an order parameter of the $U(1)'$ symmetry breaking, yielding a hierarchy between the $Z'$ mass and the EW scale. Below we will be working in this context. For simplicity we further turn off the couplings between the $\{ \bf H_d, H_u, S\}$ sector and the secluded-$U(1)'$ breaking sector in the superpotential and soft SUSY-breaking terms. Then the influence of the $U(1)'$ gauge symmetry on the Higgs potential at tree level enters via D-terms only. The mixing between the $\{ \bf H_d, H_u, S\}$ sector and the secluded-$U(1)'$ breaking sector can be safely neglected. Without loss of generality, we interpret the 750 GeV resonance as an $S$-like scalar (the analysis is similar in most cases if the $S$-like pseudoscalar is the candidate for the 750 GeV resonance).

With these assumptions, the superpotential and soft supersymmetry-breaking terms of the Higgs sector are given by 
\beq
\begin{split}
{\bf W_H}& = \lambda {\bf S H_u H_d} + \Delta {\bf W_{H} (S, S_i)}   \\
V_{\rm soft}^H &=  - A_\lambda \lambda  S H_u H_d + {\rm h.c.}  \\ & \quad + m_{H_d}^2 |H_d|^2 + m_{H_u}^2 |H_u|^2 + m_{S}^2 |S|^2 \\
& \quad + f (S, S_i)  
\end{split}
\eeq
yielding a tree-level mass matrix for the CP-even Higgs sector in the basis $\{H_d^{r} \equiv Re(H_d), H_u^{r}, S^{r}\}$
\begin{eqnarray}
\mathcal{M}_{\rm Even} = \left(\begin{matrix} \kappa_{H_d}^2  & \kappa_{H_d, H_u} &
\kappa_{H_d, S} \cr \kappa_{H_d, H_u} & \kappa_{H_u}^2  &
\kappa_{H_u, S}  \cr \kappa_{H_d, S}  & \kappa_{H_u, S}  &
\kappa_S^2  \end{matrix}\right) ,\,
\end{eqnarray}
where
\hspace{-3mm}
\begin{eqnarray}
\begin{split}
\kappa_{H_i}^2 &= 2 \left( {{G^2} \over {4}} + g_{Z'}^2 q_{H_i}^2
\right) v_i^2  +\lambda^2 (v_j^2 +v_S^2)\\
&\quad + m_{H_i}^2+|\epsilon_{i j}|{{G^2}\over 4}
(v_i^2-v_j^2) +g_{Z'}^2 q_{H_i} \Delta  \\
\kappa_S^2 &= 2 g_{Z'}^2 q_S^2 v_S^2  +
\lambda^2 (v_d^2 +v_u^2)  +  m_{S}^2 + g_{Z'}^2 q_{S} \Delta  \\
\kappa_{H_{d}, H_{u}} &= 2 \left( \lambda^2 - {{G^2} \over {4}} + g_{Z'}^2
q_{H_d} q_{H_u} \right) v_d v_u \\&\quad  - A_\lambda \lambda v_S  \\
\kappa_{H_i, S} &= 2 \left( \lambda^2 + g_{Z'}^2 q_{H_i} q_S \right) v_i v_S
 - |\epsilon_{i j}| A_\lambda \lambda v_j 
\end{split}
\end{eqnarray}
with 
\begin{eqnarray}
\Delta \equiv q_S v_S^2 + q_{H_d} v_d^2 + q_{H_u} v_u^2 + \sum_{i=1}^3 q_{S_i} v_{S_i}^2
~.~\,
\end{eqnarray}

Given the 125~GeV Higgs that has been observed at the LHC is SM-like, we naturally consider the alignment limit with negligibly small tree-level mixing between $S$ and $H_d$, $H_u$. Such a limit can be achieved by an assumption, e.g.,  
\beq
\label{eq:condition}
\tan\beta =1, \ q_{H_u}=q_{H_d},  \  A_\lambda = \frac{2}{\lambda} (\lambda^2 + q_{H_u} q_S g_{Z'}^2) v_S
\eeq
Next, using the minimization conditions we replace $m_{H_d}^2$, $m_{H_u}^2$ and $m_S^2$ with $v_d$, $v_u$ and $v_S$. Then the tree-level mass eigenvalues are solved to be 
\beq
 \label{mass}
\begin{split}
m_{h}^2 &=(2 g_{Z'}^2 + \lambda^2) v^2 ,  \\
m_{H}^2 &=\frac{G^2}{2} v^2 - 8 g_{Z'}^2 v_S^2 - \lambda^2 v^2 + 
   4\lambda^2 v_S^2,  \\ 
m_{Y}^2 &= -2 g_{Z'}^2 v^2 +8 g_{Z'}^2 v_S^2 +\lambda^2v^2. 
\end{split}
\eeq
Here $h$ is the SM-like Higgs, which receives tree-level contributions to its mass from both F-terms and D-terms. $H$ is the MSSM-like CP-even Higgs and $Y$ is the singlet-like one. As discussed above, the secluded $U(1)'$-breaking sector mainly serves as an order parameter of the $U(1)'$ symmetry breaking. Indeed, the $\Delta$-dependent terms which involve $v_{S_i}$ are gone in Eq.(\ref{mass}). 

In this context the coupling $g_{YHh}$ and $g_{Yhh}$ generically vanish at tree level. We do not expect $Y$ be produced via $H\to Y h$, even if  $H$ might be heavier than $Y$. Interestingly, the coupling 
\beq
\begin{split}
g_{YHH} &= -8 g_{Z'}^2 v_S  + 4 \lambda^2 v_S + 2 A_\lambda \lambda \\
&\quad \approx -2 \frac{m_Y^2}{v_S} + 8 \lambda^2 v_S
\end{split}
\eeq
could be sizable. The three terms in the first line represent contributions from D-terms, F-terms, and softly SUSY-breaking terms, respectively. To accomplish the second line, the equality relation~(\ref{eq:condition}) has been taken and $m_Y^2 \approx 8 g_{Z'}^2 v_S^2$ has been assumed.

The superpotential and softly supersymmetry-breaking terms for describing the interaction between $S$ and the vector-like fermions and their superpartners  are 
\beq
\begin{split}
\hspace{10mm} {\bf W_F}& = \kappa {\bf S F_V F_V^c}    \\
\hspace{10mm} V_{\rm soft}^F &=  A_\kappa \kappa  S\tilde  F_V  \tilde F_V^c + {\rm h.c.} 
\end{split}
\eeq
Then the effective couplings between $Y$ and di-gluons and di-photons are yielded by
\begin{eqnarray}
g_{Y F_VF_V^c} \!\!\!\! &=& \!\!\!\! \kappa  \label{eq:VLQcoup}\\ 
g_{Y \tilde F_V \tilde F_V^*} = g_{Y \tilde F_V^c \tilde F_V^{c*}} \!\!\!\! &=& \!\!\!\! 2\kappa v_S \label{eq:SVLQcoup}
\end{eqnarray}
if there is no mixing between $\tilde F_V$ and $\tilde F_V^c$. This indicates that $\kappa$ is a crucial parameter in determining the production and decay patterns of the $Y$ resonance. Additionally, the coupling between $Y$ and di-$Z'$ bosons is given by
\beq
\hspace{10mm} g_{YZ'Z'} = 4\sqrt{2} g_{Z'}^2  v_S 
\eeq
Using these outcomes, it is straightforward to calculate the signal rates in various scenarios. 

\section{Benchmark Scenarios}

The PQ-type $U(1)'$ with the MSSM matter content only is anomalous. To cancel the mixing anomalies between the SM gauge symmetries and $U(1)'$,  
vector-like color triplets and EW multiplets carrying $U(1)'$ charges are typically required, no matter in the GUT (see, e.g.,~\cite{Kang:2004pp, Kang:2009rd}) or in the non-GUT (see, e.g.,~\cite{An:2012vp}) frameworks.  Such a construction automatically warrants that no new anomalies are introduced for the SM gauge symmetries.  As an illustration, we will work on the model built in~\cite{An:2012vp} with a mild modification.  The particle spectrum is listed in Table~\ref{table1}, which includes three classes of supermultiplets: the ones defined in the MSSM, the exotic ones charged under the SM gauge symmetries and their signlets. All of these supermultiplets carry the $U(1)'$ gauge charges. Note that in Table~\ref{table1} ${\bf S_{1,2}}$ and ${\bf S_{1,2}^c}$ are not required by anomaly cancellation. We introduce them by hand, in order to generate a mass term for $\{{\bf  X_1, X_1^c}\}$ and $\{{\bf  S_3, S_3^c}\}$, respectively. 

\begin{table}[t]
\caption{Particle content in an anomaly-free supersymmetric $U(1)'$ model, with their $SU(3)\times SU(2)_L\times U(1)_Y\times U(1)'$ gauge charges.} 
\vspace{7 mm}
\label{table1}
\begin{center}
\begin{tabular}{|c|c|c|c|}
\hline
  & Gauge charges  &  &  Gauge charges \\
\hline
\hline 
${\bf L_i}$ & (1; 2; $-1/2$; $1/2$) &
${\bf Q_i}$ & (3; 2; 1/6; $1/2$) \\
\hline
${\bf \bar N_i}$ & (1; 1; 0; $1/2$) & ${\bf \bar u_i}$ & ($\bar 3$; 1; $-2/3$;  $1/2$)\\
\hline
${\bf \bar e_i}$ & (1; 1; 1; $1/2$) & ${\bf \bar d_i}$ & ($\bar 3$; 1; 1/3;  $1/2$)\\
\hline 
${\bf H_d}$ & (1; 2; $-1/2$; $-1$) & ${\bf H_u}$ & (1; 2; $1/2$; $-1$) \\
\hline
\hline
${\bf T_1}$ &  (3; 1; $2/3$; $-1$) & $ {\bf T_1^c}$ &  ($\bar 3$; 1; $-2/3$; $-1$)  \\
\hline
${\bf T_2}$ &  (3; 1; $-1/3$; $-1$) & $ {\bf T_2^c}$ &  ($\bar 3$; 1; $1/3$; $-1$)  \\
\hline
${\bf T_3}$ &  (3; 1; $-1/3$; $-1$) & $  {\bf T_3^c}$ &  ($\bar 3$; 1; $1/3$; $-1$)  \\
\hline
${\bf D_1}$ &  (1; 2; $1/2$; $-1$) & $  {\bf D_1^c}$ &  (1; 2; $-1/2$; $-1$)  \\
\hline
${\bf D_2}$ &  (1; 2; $1/2$; $-1$) & $  {\bf D_2^c}$ &  (1; 2; $-1/2$; $-1$)  \\
\hline
$ {\bf  X_1}$ & (1; 1; 1; $2$) & $ {\bf X_1^c}$ & (1; 1; -1; $2$) \\
\hline
$ {\bf  X_2}$ & (1; 1; 1; $-1$) & $ {\bf X_2^c}$ & (1; 1; -1; $-1$) \\
\hline
\hline
${\bf S_1}$ & (1; 1; 0; $4$) &${\bf S_1^c}$ & (1; 1; 0; $-4$)  \\
\hline
${\bf S_2}$ & (1; 1; 0; $2$) &${\bf S_2^c}$ & (1; 1; 0; $-2$)  \\
\hline
${\bf S_3}$ & (1; 1; 0; $1$) &${\bf S_3^c}$ & (1; 1; 0; $1$)  \\
\hline
${\bf S}$ & (1; 1; 0; $2$) & &   \\
\hline
\end{tabular}
\end{center}
\end{table}

\begin{figure}[ht]
\begin{center}
\hspace*{3mm}\includegraphics[width=0.4\textwidth]{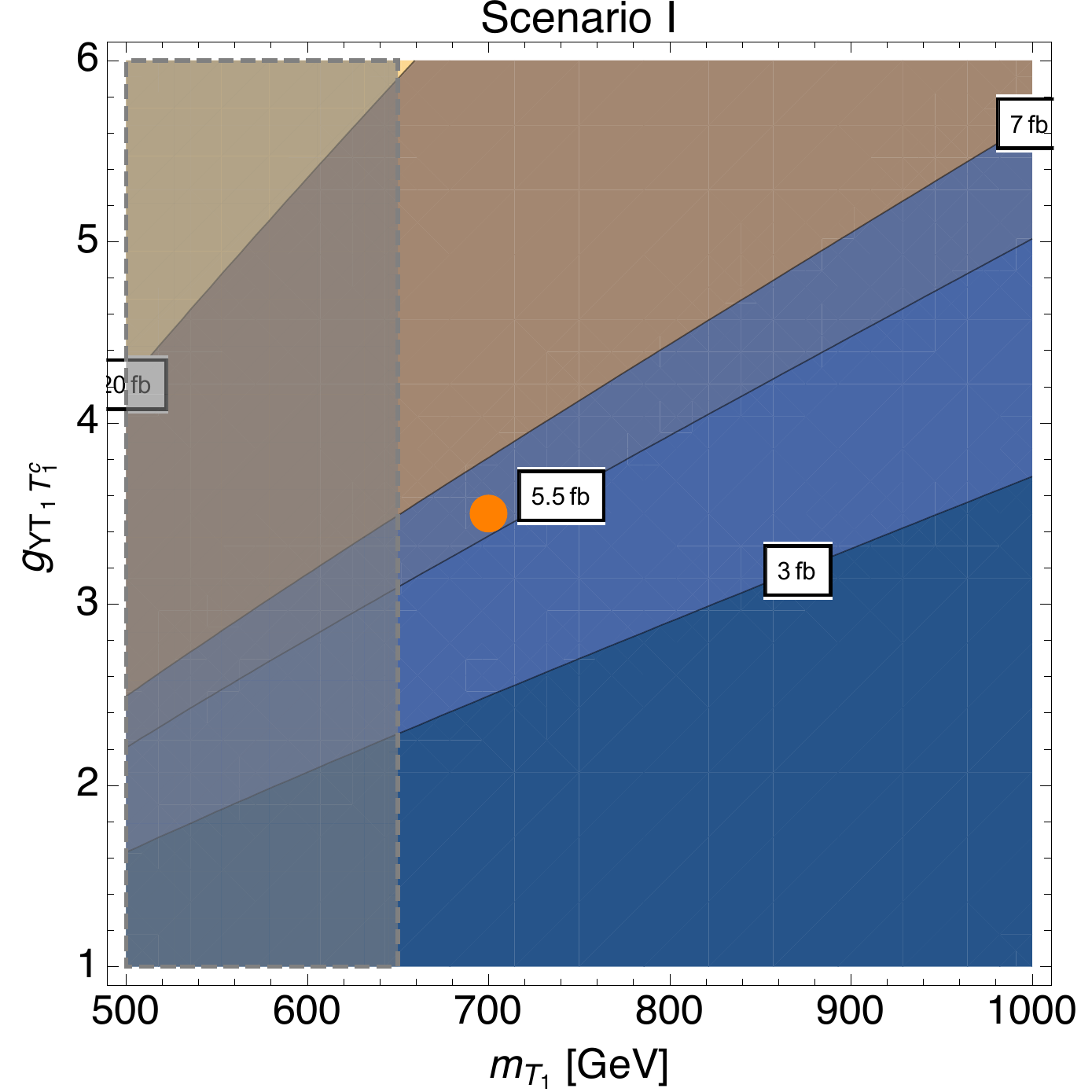}\\[7pt]
\hspace*{3mm}\includegraphics[width=0.4\textwidth]{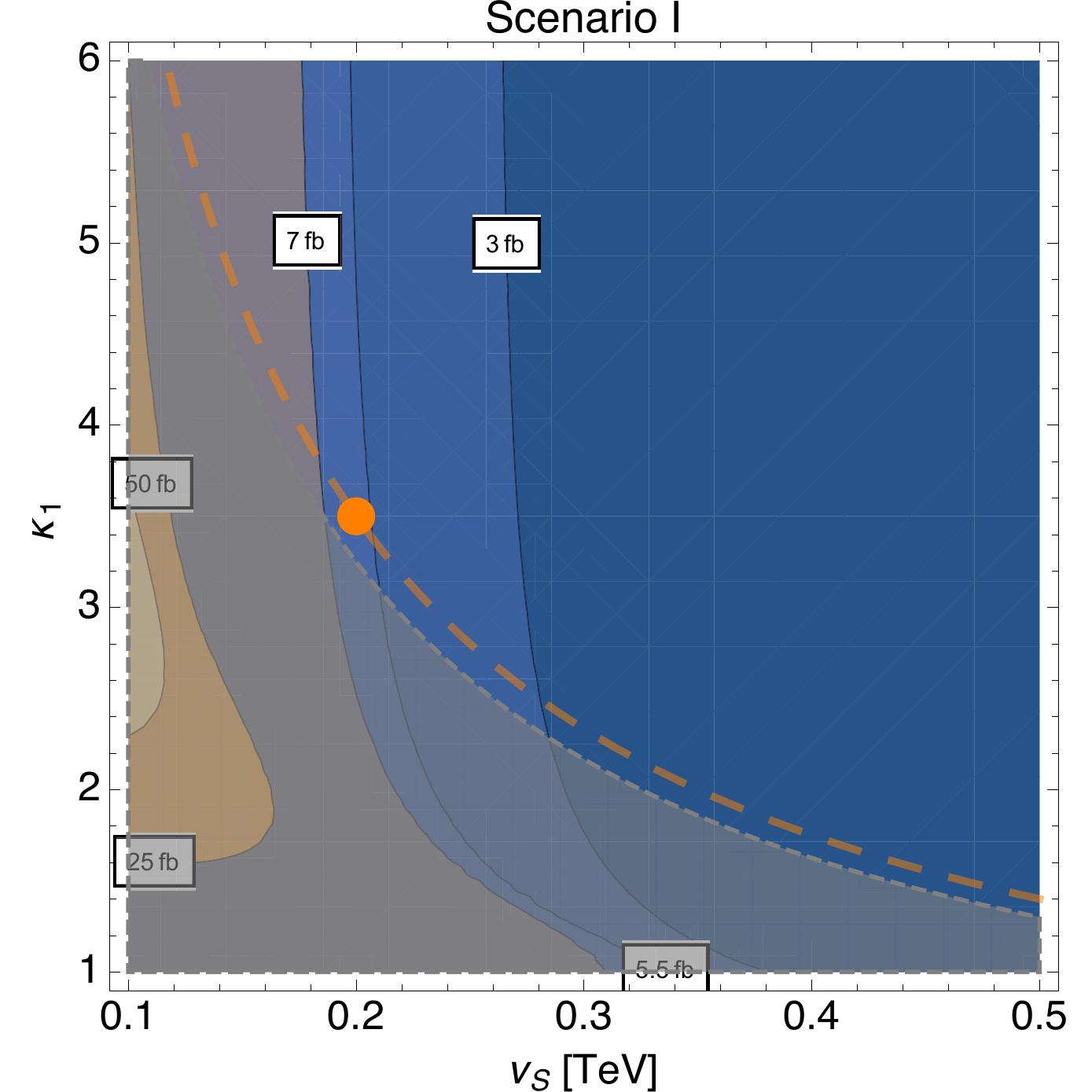}
\caption{Contours of $\s(pp\to Y\to \g\g)$ in Scenario~I, with the shaded region excluded by the requirement of $m_{T_1} > 650$ GeV (see discussions in the text). The orange dot is a benchmark point presented in Table~\ref{BMPtab}. In the bottom panel, the orange dashed curve is defined with $m_{T_1}=m_{T_1^c} = \kappa_1 v_S = 700$ GeV. Here the decay of $Y\to T_1 T_1^c$ is kinematically turned off.}
\label{contourplot1}
\end{center}
\vspace{1mm}
\end{figure}

\begin{figure}[ht]
\begin{center}
\hspace*{3mm}\includegraphics[width=0.4\textwidth]{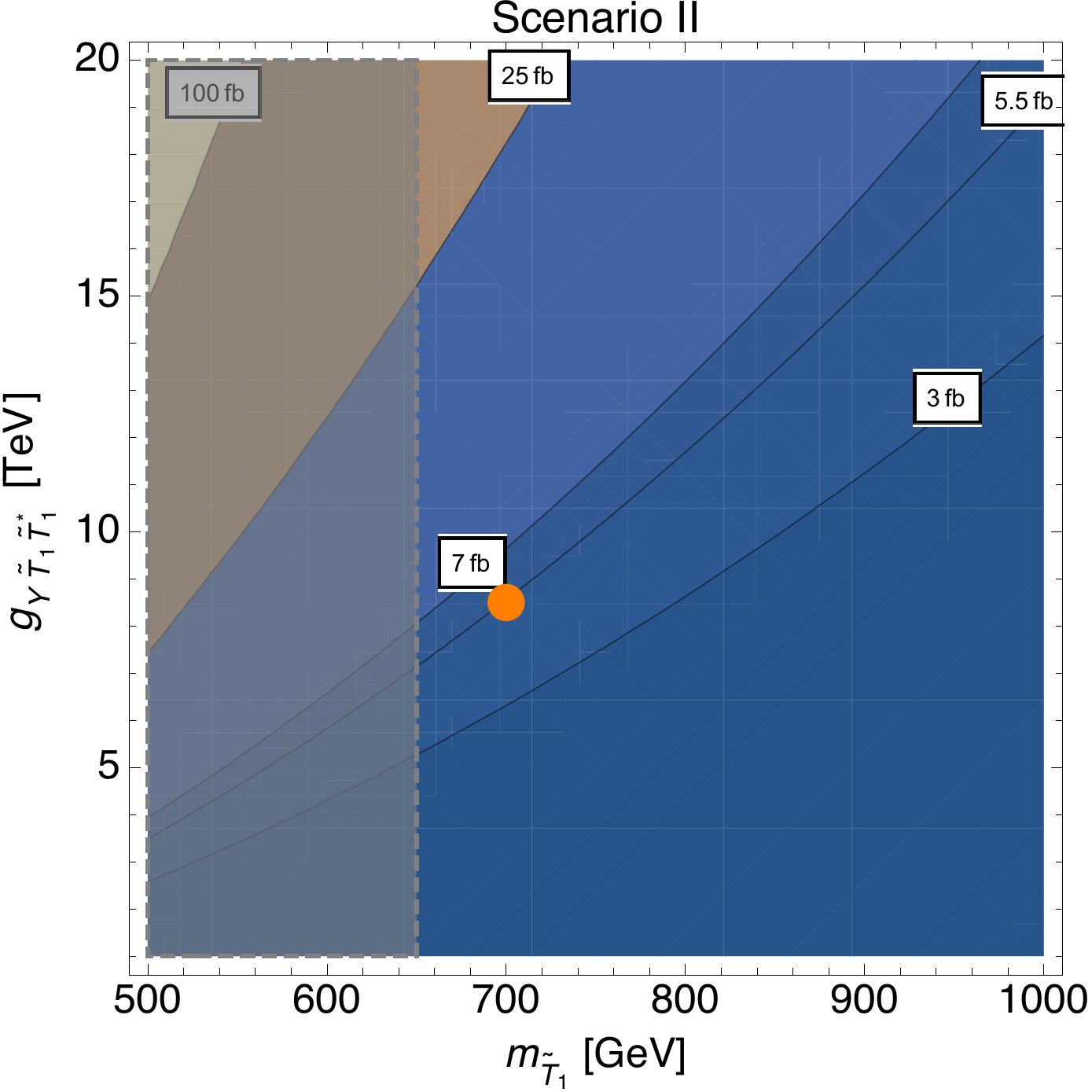}\\[7pt]
\hspace*{3mm}\includegraphics[width=0.385\textwidth]{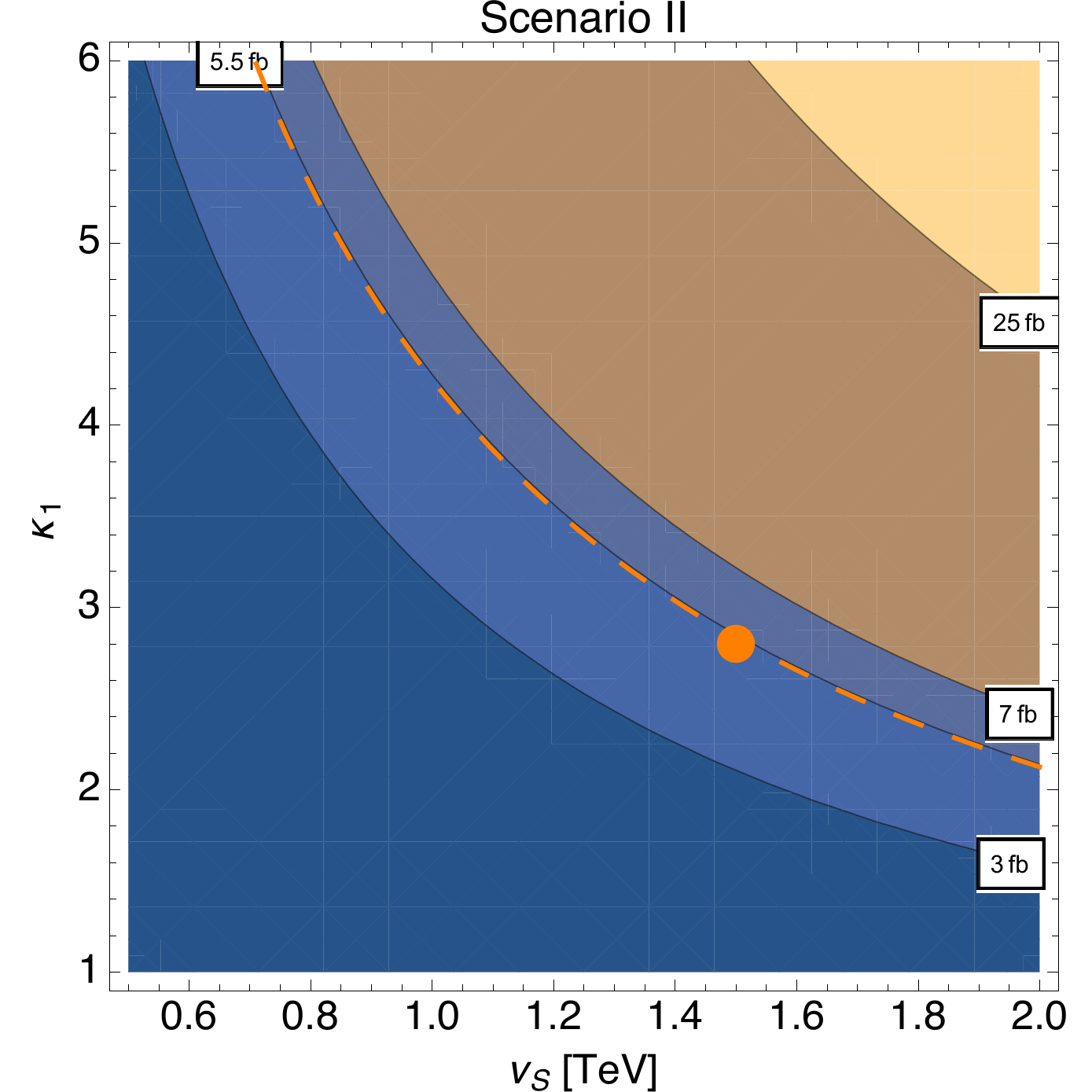}
\caption{Contours of $\s(pp\to Y\to \g\g)$ in Scenario~II, with the shaded region excluded by the requirement of $m_{\tilde T_1} > 650$ GeV (see discussions in the text). The orange dot is a benchmark point presented in Table~\ref{BMPtab}. 
The region above the red line (top panel) has been excluded by the di-jet searches at the LHC. 
In the bottom panel, $m_{\tilde T_1}=m_{T_1^c}=700$ GeV is assumed. The orange dashed curve is defined with $g_{Y \tilde T_1 \tilde T_1^{*}}=g_{Y \tilde T_1^c \tilde T_1^{c*}} = 8.5$ TeV.
}
\label{contourplot2}
\end{center}
\vspace{1mm}
\end{figure}

Some relevant terms in the superpotential, ${\bf W_{\rm F}}$, for the second class of supermultiplets are given by 
\begin{equation*}
\begin{split}
{\bf W_{\rm F}} &\sim \kappa_i {\bf S T_i T_i^c} +\gamma_r {\bf S D_rD_r^c} + \zeta_ 1{\bf S_1^c X_1X_1^c} + \zeta_2 {\bf S X_2X_2^c} \\
 &+ \kappa_{ij1} {\bf \bar N_i \bar u_j T_1} +  \kappa_{ijk} {\bf \bar N_i \bar d_j T_{k=2,3}}  +   \kappa'_{ijk} {\bf \bar e_i \bar u_j T_{k=2,3}}   \\
& + \gamma_{ijr} {\bf  L_i \bar N_j  D_r } + \gamma'_{ijr} {\bf  Q_i \bar u_j  D_r } \\
& + \gamma_{ijr}^c {\bf  L_i \bar e_j  D_r^c } + \gamma_{ijr}^{'c} {\bf  Q_i \bar d_j  D_r^c } \\
&+ \zeta_{rs1} {\bf D_r^c D_s^c X_1}  +\zeta_{rs1}^c {\bf D_r D_s X_1^c} \\
& +\zeta_{ij2} {\bf \bar N_i \bar e_j X_2^c} + ... ...
\end{split}
\end{equation*}
Here $i, j$ run over 1, 2 and 3, and $r, s$ run over 1 and 2.  After the $U(1)'$ symmetry is broken, $\{{ \bf T_i, T_i^c}\}$, $\{{\bf D_r, D_r^c}\}$ and $\{{\bf X_r, X_r^c}\}$ obtain vector-like masses by coupling with the supermultiplet ${\bf S}$ (or ${\bf S_1^c}$). The supermultiplet ${\bf S_2}$ may also contribute if its scalar field obtains a non-zero VEV and meanwhile, couples with these vector-like supermultiplets. An important feature is that these vector-like supermultiplets can decay into the SM particles via proper interactions listed above, avoiding the overproduction in the Universe. Some searches related to these exotics have been pursued at the LHC. The current bounds are dependent on the SM gauge charges carried by these exotics as well as their couplings with the SM particles. $\{{ \bf T_i, T_i^c}\}$, in essence, are supermultiplets of leptoquarks. The LHC searches for leptoquarks have been pursued at both the ATLAS and the CMS~\cite{Aad:2015caa,Khachatryan:2015vaa,Khachatryan:2015bsa}. If they exclusively decay into the SM fermions of third-generation, the data collected at Run-1 set a lower bound for their mass $\sim 600-650$~GeV~\cite{Aad:2015caa,Khachatryan:2015bsa}. As for their fermionic superpartners, if they also exclusively decay into the same final states with one additional lightest superparticle which is neutral, comparable mass bounds are expected. Whereas, $\{{\bf D_r, D_r^c}\}$ and $\{{\bf X_r, X_r^c}\}$ are colorless supermultiplets. As a matter of fact, the gauge charges carried by $\{{\bf D_r, D_r^c}\}$ are exactly the same as the ones carried by  $\{{\bf H_u, H_d}\}$. Thus we do not expect a strong bound for their mass in a general context. In this article, we simply assume a universal lower bound of 650 GeV for the mass of both leptoquarks and their superpartners, as indicated in Fig.~\ref{contourplot1} and Fig.~\ref{contourplot2}.  

In the following we discuss three representative scenarios that could yield the di-photon signal for the 750 GeV resonance. In all cases we assume no mixing between the Higgs doublets and the singlet $S$, by demanding the relation~(\ref{eq:condition}). The pure singlet scalar $S$ accounts for the 750~GeV resonance. 

{\bf Scenario I:}  VLQ-mediated $pp \to Y \to \gamma \gamma$. The candidates for the VLQ mediator are the fermion component of the superfields $\{{\bf T_i, T_i^c  }\}$. 

{\bf Scenario II:} SVLQ-mediated $pp \to Y \to \gamma \gamma$. The candidates for the SVLQ mediator are the scalar component of the superfields $\{{\bf T_i, T_i^c  }\}$.  

{\bf Scenario III:} $H/A$-mediated production mode $pp \to H^*/A^*\to Y H/h$, with the production of $H^*/A^*$ dominated by gluon fusion and the decay $Y \to \gamma\gamma$ mediated by vector-like leptons (VLL) or scalar VLL (SVLL). The superfields $\{{\bf D_r, D_r^c}\}$ and $\{{\bf X_2, X_2^c}\}$ provide the VLL or SVLL mediators for the $Y \to\g\g$ decay. $\{{\bf X_1, X_1^c}\}$ does not couple with ${\bf S}$ directly and hence can not serve for this purpose. Though a three-body process $pp \to H \to Y H^*$ is also likely, if $m_{H} > m_Y $, the decay width of $H \to Y H^*$ is too small, compared with that of $H\to t\bar t$, to make this process significant. So we will focus on the former case.

We present in Fig.~\ref{contourplot1} the contours of the di-photon signal rate $\s(gg\to Y\to \g\g)$ in Scenario~I at the $m_{T_1} - g_{Y T_1 T_1^c}$ plane (upper panel) and at the $v_S - \kappa_1$ plane (bottom panel). The VLQ mass $m_{T_1}=m_{T_1^c} =\kappa_1 v_S$ is assumed here. The corresponding results in Scenario~II are shown at the $m_{\tilde T_1} - g_{Y{\tilde T_1} {\tilde T_1^*}}$ plane (upper panel) and at the $v_S - \kappa_1$ plane (bottom panel) in Fig.~\ref{contourplot2}, with an assumption of $m_{\tilde T_1} = m_{\tilde T_1^c}$\footnote{Unlike the VLQ, the SVLQ may receive contributions to its mass not only from F-terms, but also from D-terms and softly supersymmetry-breaking terms. For sake of brevity, we are not attempting to explicitly write out the full expression.} and  $g_{Y \tilde T_1 \tilde T_1^{*}}=g_{Y \tilde T_1^c \tilde T_1^{c*}}$. Alhough all of the three pairs of VLQ supermultiplets can contribute in both scenarios, we include the contributions from $\{{\bf T_1, T_1^c}\}$ only, assuming the other VLQ supermultiplets be decoupled. 

Remarkably, the desired signal rate $\s(pp\to Y\to \g\g)$ of order $\sim \mathcal{O}(10)$~fb can be achieved in both Scenario I and II, and hence fits well both the ATLAS and CMS signals at 13 TeV LHC. The signal rate in Scenario~II shows a strong dependence on the mediator mass, simply because the effective di-photon coupling is proportional to a factor of $1/m_{\tilde T_1}^{2}$ or $1/m_{\tilde T_1^c}^{2}$ other than a loop-function.  
Whereas, such a strong dependence on the mediator mass is absent in Scenario~I (mediated by a fermion VLQ).
Compared to Scenario I, a larger $\kappa_1$ is typically required to yield a comparable signal rate in Scenario II. This partially results from the fact that the ratio between a scalar-loop function and a fermion-loop function asymptotically approaches a value ${\mathcal{A}_f / \mathcal{A}_s} \simeq 4$ as the mediator mass increases. The di-jet search at the Run-1 have a marginal impact on eliminating the parameter space. The projection of the Run-1 limit to the 13 TeV Run-2 yields a bound $\s(gg\to Y \to gg)<100$~pb~\cite{Aad:2014aqa}. Since the partial width of di-gluon decay is only $\sim400$ times larger than that of the di-photon decay in our case, the di-jet bound is weak in both Scenario~I and Scenario~II.

\begin{figure}[t]
\begin{center}
\hspace*{3mm}\includegraphics[width=0.4\textwidth]{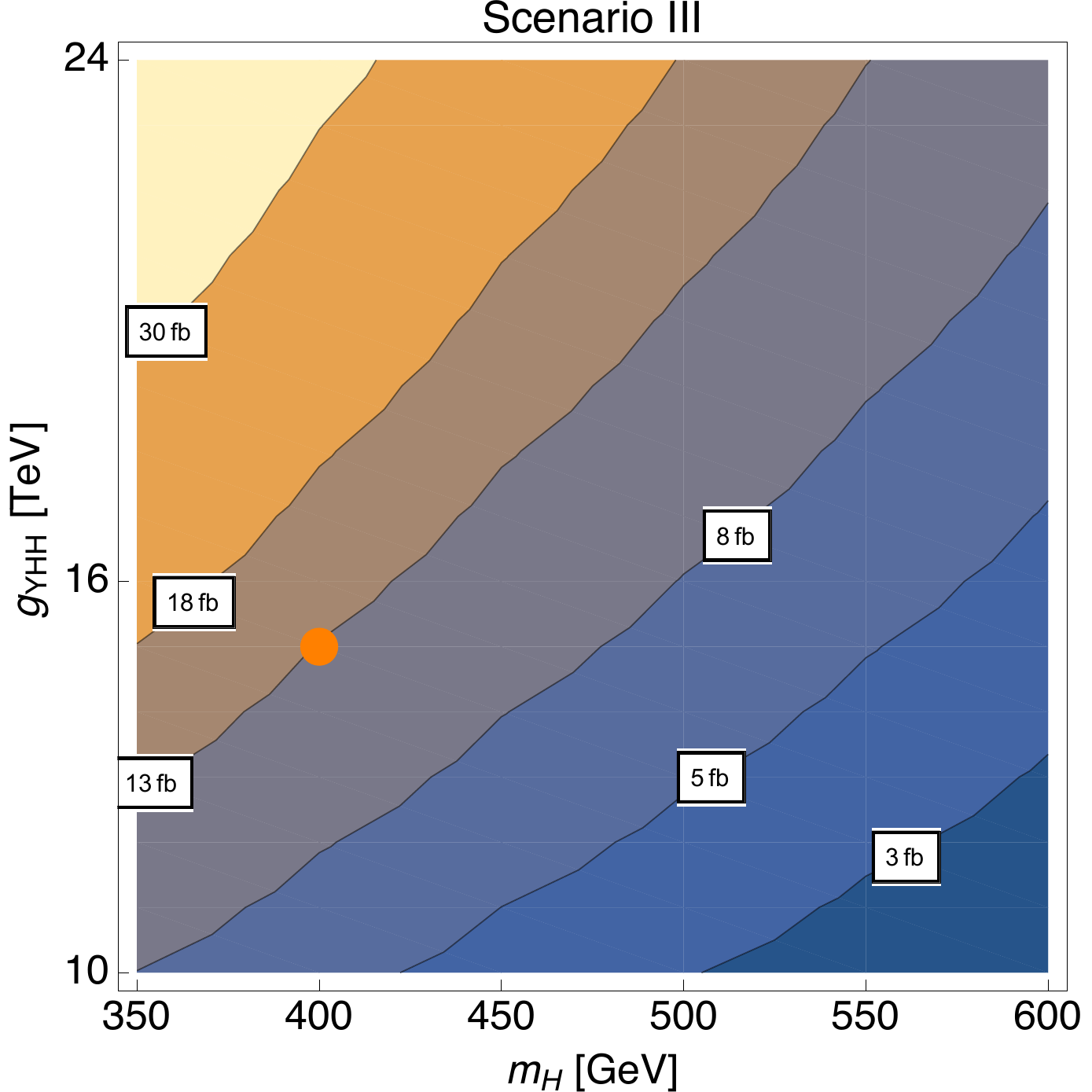}\\[7pt]
\hspace*{3mm}\includegraphics[width=0.4\textwidth]{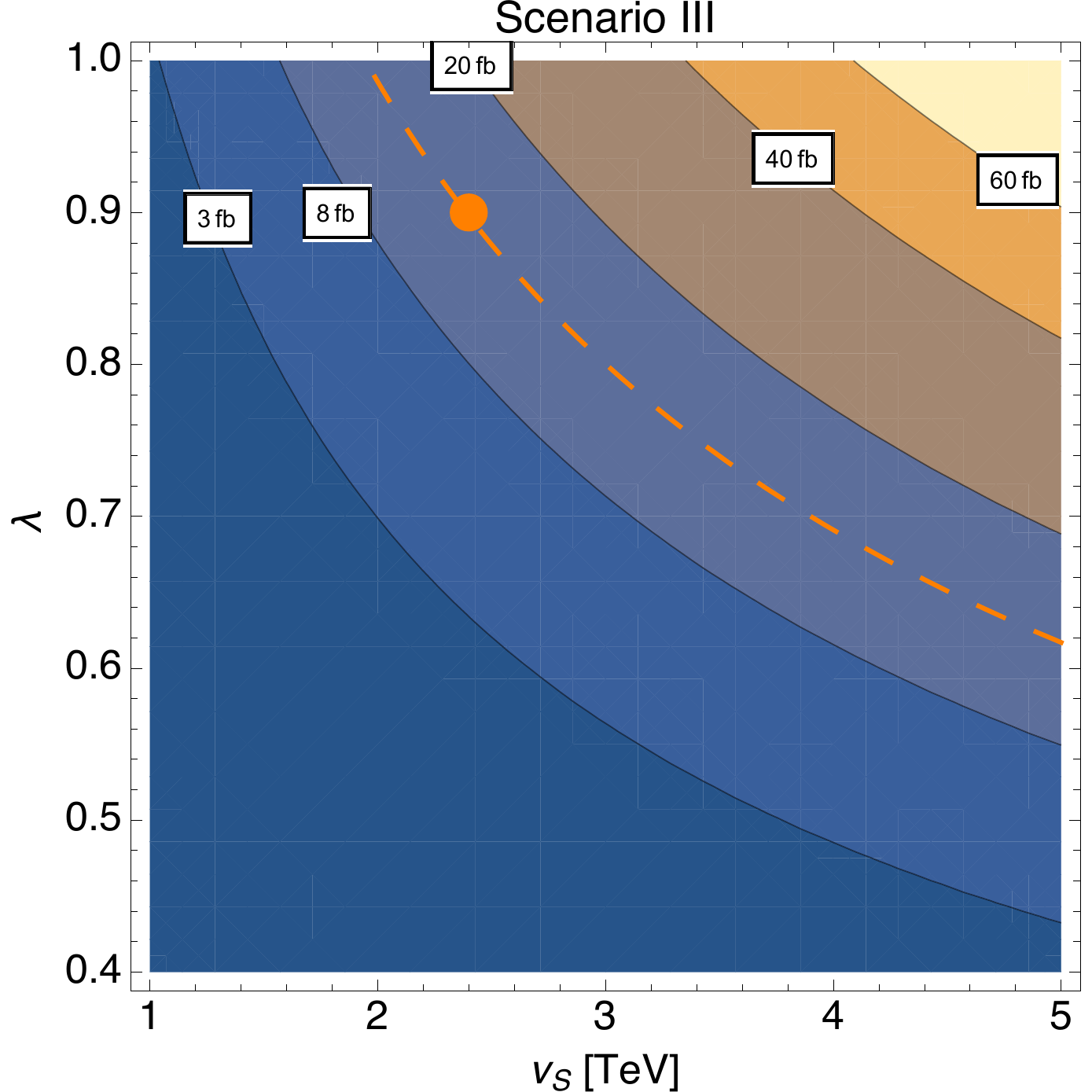}
\caption{Contours of $\s(pp\to H^*\to YH)$ in Scenario~III. The orange dot is a benchmark point presented in Table~\ref{BMPtab}. In the bottom panel, the Higgs mass is fixed to be $m_H=400$ GeV.  The orange dashed curve is defined with $g_{YHH} = 15$ TeV, yielding $\s(pp\to H^*\to YH)=13.0$ fb.}
\label{contourplot3}
\end{center}
\vspace{1mm}
\end{figure}

The results of Scenario~III are displayed in Fig.~\ref{contourplot3}. Clearly, there exists parameter space where a cross section of $\mathcal O(10)$~fb or above for the $Y$ production in association with a nonstandard Higgs boson $H$ is obtainable. If the di-photon coupling of $Y$ is mediated by VLL or SVLL only, such as the ones in $\{ {\bf X_2, X_2^c} \}$, the branching ratio of its decay into di-photon can be as large as above 50\%, due to the absence of the $Y\to WW, gg$ decays. As such, the recently observed signal rate on $Y$ at both the ATLAS and the CMS could be explained. 

In addition, the 750 GeV resonance could be generated via the $Z'$-mediated VBF production $pp \to Y qq'$, with the decay $Y \to \gamma\gamma$ mediated by VLL or SVLL in the loop. However,  in this scenario a large signal rate favors a light $Z'$ and a large $U(1)'$ gauge coupling. We find that $\s(pp\to Y\to \gamma\gamma)\sim 10$~fb can only result from a $Z'$ boson which might have been excluded by the Drell-Yan process search at the LHC~\cite{CMS:2015nhc}. Thus, even if this scenario works, we believe it will be extremely fine-tuning.

Finally, we collect in Table~\ref{BMPtab} three benchmarks points (BMPs), representing the three benchmark scenarios discussed above, respectively. They are all able to produce the di-photon signal rates of the order observed. Should future analyses confirm this signal, it will be of interest to fully test these scenarios using complementary decay channels in experiment. Using the Run-1 data, both the ATLAS and the CMS have analyzed the upper limits on the production cross section of new particle $Y$ at 750~GeV which decay into various final states including $ZZ$, $WW$, $hh$, $Zh$, $Z\g$, $\tau\tau$ and $t\bar{t}$. A crude estimation of these upper limits, rescaled by  the upper limit of $\s(gg\to Y\to \g\g)$ of an order $\sim \mathcal O(1)$ fb,  has been offered in~\cite{Ellis:2015oso}:   $R(ZZ,WW,hh,Zh,Z\g,\tau\tau,t\bar{t}) \lesssim (23,46,41,20,7,20,600)$, with $R(XX) = {\rm BR}(Y\to XX)$ $/{\rm BR}(Y\to \g\g)$. The $Z\g$ search thus may play a crucial role in this regard. Additionally, these scenarios could be probed by searching for the mediators of the di-photon couplings. We can also probe Scenario III by searching for non-standard Higgs bosons via its decay into $t\bar t$. At last, we would point out - the couplings $\kappa_1$ and $\lambda$ of $\sim \mathcal O(1)$ may yield a Laudau problem. Particularly in Scenario I and II, $\kappa_1$ may hit its Landau pole at a energy scale as low as 10 TeV. We refer to~\cite{Gu:2015lxj} for more discussions in this regard. Turning on more colored mediators, such as the component field in $\{\bf T_{2,3}, T_{2,3}^c \}$ in this model, in the di-photon and di-gluon loops doesn't help much, since their relatively small hypercharges or electric charges further suppress the branching ration of the di-photon decay. Alternatively, one may turn on color-neutral mediators simultaneously, which may greatly enhances the branching ratio of the di-photon decay. In this case, the $kappa$ value required for generating the signal could be suppressed to be below $\mathcal O(1)$.

\begin{table}[t]
\caption{Benchmark points  in each scenario. These points are marked in orange dots in Figs.~\ref{contourplot1} - \ref{contourplot3}. }
\label{BMPtab}
\vspace{10mm}
\begin{center}
\begin{tabular}{|c|c|c|c|c|c|}
\hline
BMP & $\kappa_1$ & $v_S$ & $M_{F_V,\tilde F_V}$ & $\s(gg \to Y \to \g\g)$ \cr
\hline
I  & 3.5 &   0.2 TeV & 0.7 TeV &  5.9 fb \cr 
II  & 2.8 &   1.5 TeV & 0.7 TeV &  5.4 fb\cr 
\hline
\end{tabular}

\begin{tabular}{|c|c|c|c|c|c|}
\hline
BMP &  $\lambda$ & $v_S$  & $M_H$ & $\s(gg \to Y \to \g\g)$  \cr
\hline
III  & 0.9 & 2.5 TeV & 0.4 TeV &  13.0  fb\cr 
\hline

\end{tabular}
\end{center}
\label{default}
\end{table}%

\section{Conclusions}

Motivated by the recent observation of a potential 750 GeV di-photon resonance at the $\sqrt{s}=13$~TeV LHC, we studied its implications for low-energy supersymmetric theories. Though it is generically difficult to explain the observed signal rate in the MSSM and in the NMSSM, due to the lack of an efficient mechanism to generate the observed signal rate, we pointed out that such a resonance can be accommodated in the MSSM-extensions with a PQ-type $U(1)'$ gauge symmetry. For illustration, we explore several benchmark scenarios in which the di-photon resonance is produced either via (scalar) VLQ-mediated gluon fusion process or in association with a heavy Higgs scalar. 
Additionally, we studied the possibility of $Z'$- mediated vector boson fusion production and $Z'$-associated production: $pp \to Y qq'$. Compared to the other scenarios, this possibility is constrained by the current experimental bounds much more strongly. 

Though our studies were focused on a CP-even candidate resonance, the discussions can be extended to a CP-odd candidate resonance, with a couple of exceptions. First, the $Z'$-mediated mechanism can not be applied to the latter case, unless CP-violation is allowed. Additionally, in this context a CP-odd singlet prefers to couple with a MSSM-like CP-odd Higgs $A$ and a SM-like Higgs $h$, instead of the $A$ plus a heavier Higgs $H$. Therefore, we may well expect a significant production of the resonance in association with a SM-like Higgs $h$, mediated by an off-shell $A$.

As a candidate theory to explain the potential 750 GeV di-photon resonance, this class of supersymmetric models have three main advantages.  
First, the singlet supermultiplet (the one providing the candidate of the 750 GeV resonance) is generically required for dynamically generating an effective $\mu$ parameter. Second, the vector-like supermultiplets are generically required by anomaly cancellation of the whole set of gauge symmetries, unless some Chern-Simons term is properly incorporated. Third, several mechanisms work well in this context, enabling the 750 GeV resonance to be accommodated in multiple ways, if it does exist. All of them do not require a mixing between the singlet and the Higgs doublets  in the 750 GeV resonance.  This certainly broadens our understanding on both the potential 750 GeV resonance and the low-energy scale supersymmetric theories. 

In spite of these advantages, the resonance exclusively decays into $gg$, $\gamma\gamma$, $Z\gamma$ and $ZZ$ in Scenarios~I and II and into $\gamma\gamma$, $Z\gamma$ and $ZZ$ in Scenario~III. This yields a partial width of $Y\to \gamma \gamma$ typically narrow, in comparison to $\sim 45$~GeV which is indicated by the ATLAS measurement~\cite{ATLAS:2015}. This effect may not cause a decay length longer than sub-millimeter, e.g., by raising the coupling value between $Y$ and $\{ {\bf X_2, X_2^c} \}$ in Scenario~III. Nonetheless, it is necessary to think about how to reconcile this discrepancy. Given the LHC may release more precise information on the properties of this resonance soon,  we leave a further study in this regard to future work.


\begin{center}
{\bf Acknowledgments}
\end{center}  

\vspace*{-5pt}

YJ is supported by the Villum Foundation. YYL is supported by the the Hong Kong PhD Fellowship Scheme (HKPFS).  TL is supported by the Collaborative Research Fund (CRF) HUKST4/CRF/13G and the General Research Fund (GRF) 16304315. Both the HKPFS and the CRF, GRF grants are issued by the Research Grants Council of Hong Kong S.A.R..


\end{document}